\documentstyle[twoside,fleqn,espcrc2]{article}

\newcommand{\epe}{\epsilon^\prime/\epsilon}
\newcommand{\be}{\begin{equation}}
\newcommand{\ee}{\end{equation}}
\newcommand{\bea}{\begin{eqnarray}}
\newcommand{\eea}{\end{eqnarray}}

\newcommand{\etal}{{\it et al}.}

\title{Lattice Extraction of $\epsilon^\prime/\epsilon$ using (PQ) Chiral Perturbation Theory}

\author{Jack Laiho\address[princ]{Department of Physics, Princeton
University,         Princeton, NJ  08544.}
 }

\begin{document}

\begin{abstract}
Results at NLO in (PQ)ChPT for extracting
$\epsilon^\prime/\epsilon$ from lattice data are discussed.
 \vspace{1pc}
\end{abstract}

\maketitle



\section{Introduction}

There are renewed attempts to calculate $K\to\pi\pi$ and $\epe$ on
the lattice \cite{noaki,blum}.  Recently we have shown
\cite{laiho} that all the amplitudes of interest for the (8,1)'s
can be obtained to NLO \cite{ktwopi} in full ChPT when one uses
lattice computations from $K\to| 0 \rangle$, $K\to\pi$ and
$K\to\pi\pi$ at the two unphysical kinematics points UK1
\cite{berntwo} $\Rightarrow m_K=m_\pi$, UK2 \cite{dawson}
$\Rightarrow m_K=2m_\pi$.  In \cite{laihotwo} we extended our
previous results for the full theory; we also presented results in
the partially quenched case.  In the full theory, we showed that
none of the quantities considered in \cite{laiho} require
3-momentum insertion, i.e., it is sufficient to consider
nondegenerate quark masses in $K\to\pi$ amplitudes. We also
demonstrated how to get the low energy constants (LEC's) needed to
construct $K\to\pi\pi$ for the (8,8)'s and (8,1)'s to NLO in the
partially quenched theory. (Note that for the (8,1)'s one can go
to NLO only for the PQS implementation, as discussed below.) There
have been a few important revisions in \cite{laihotwo} since the
lattice conference.  Though our general conclusion that one can
get all of the needed combinations of NLO LEC's in the full and
partially quenched theory still holds, there are difficulties in
calculating (8,1), $K\to\pi\pi$ amplitudes at UK1 ($m_K=m_\pi$)
that must be circumvented by considering other unphysical
kinematics in order to get all the NLO constants.

There were three main additions to \cite{laihotwo} since the time
of the lattice conference.  First, we have demonstrated that the
NLO LEC's needed to construct $K\to\pi\pi$ for the (8,8)'s can be
obtained in the partially quenched case from $K\to\pi$ with
degenerate quark masses without momentum insertion and from $K\to
0$. When the number of dynamical flavors is three, the LEC's of
PQChPT are those of the full theory \cite{sharpe}.

The second addition to the paper was a discussion of the power
subtractions that are needed when one calculates $\Delta I=1/2$
amplitudes. The effects of the subtraction are considered at NLO
for the (8,8) and (8,1) amplitudes of interest.  This subtraction
makes use of the $\Theta^{(3,\overline{3})}$ operator introduced
in \cite{bern}, and is needed to perform the $K\to0$ subtraction
of $\Delta I=1/2$ amplitudes as described in \cite{blum}.  It was
discovered that the operator subtraction for the (8,1),
$K\to\pi\pi$ amplitude has a problem when the quark masses are
degenerate, and in this case, the subtraction is expected to be
quite difficult.

The third main addition to \cite{laihotwo} was an inclusion of
results at other unphysical kinematics points for (8,1),
$K\to\pi\pi$ amplitudes. These kinematics points are such that all
mesons are at rest, and the weak operator inserts energy to ensure
4-momentum conservation. We call this set of unphysical kinematics
UKX, since UK1 and UK2 are just two special cases of UKX.  If
lattice calculations at UK1 prove to be difficult or impossible,
one can use UKX as an alternative way to get all of the necessary
NLO LEC's for the (8,1)'s.

Since the time of the conference there appeared a paper by Lin,
\etal, \cite{linthree}, where they have made some important
observations regarding our attempts to obtain $K\to\pi\pi$ to NLO.
The full implications for our work are discussed below, but we
mention here that our general conclusion that one can obtain all
of the NLO LEC's in the partially quenched theory needed for
(8,1), $K\to\pi\pi$ amplitudes remains true, at least in
principle. We also review the ambiguity in embedding the partially
quenched theory in (P)QChPT as first discussed by Golterman and
Pallante \cite{goltthree}.

\section{(8,8), $K\to\pi\pi$ amplitudes with degenerate quark masses}

The following expressions are for (8,8) $K\to\pi$ amplitudes with
degenerate valence quark masses and no momentum insertion,

\bea \langle \pi^{+}|{\cal O}^{(8,8),(3/2)}|K^{+}\rangle_{ct}& =&
\frac{4\alpha_{88}}{f^{2}}+\frac{4}{f^2}[(c^r_1+c^r_2\nonumber
\\ && +4c^r_4+4c^r_5)m^2 \nonumber \\ && +2c^r_6Nm^2_{SS}], \eea

\bea \langle \pi^{+}|{\cal
O}^{(8,8),(1/2)}_{sub}|K^{+}\rangle_{ct}& =&
\frac{8\alpha_{88}}{f^{2}}+\frac{4}{f^2}[(c^r_1-c^r_2 \nonumber \\
&& -2c^r_3+8c^r_4+8c^r_5)m^2 \nonumber \\ && +4c^r_6Nm^2_{SS}].
\eea

\noindent One could, in principle, determine $\alpha_{88}$ from
either leading order term, but in practice it is safer to use the
3/2 amplitude because the 1/2 term has a power divergence, and the
subtracted amplitude could receive some residual chiral symmetry
breaking contribution unless one uses a discretization that has
exact chiral symmetry.  By varying the valence and sea quark
masses for $K\to\pi$ one can determine three linear combinations
of NLO LEC's, $c^r_6$, $c^r_1+c^r_2+4c^r_4+4c^r_5$ and
$c^r_1-c^r_2-2c^r_3+8c^r_4+8c^r_5$ from fits using Eqs (1,2), as
well as the logarithmic parts given in Appendix D of
\cite{laihotwo}.  One can verify that these combinations are
sufficient to determine the physical NLO $K\to\pi\pi$ expressions,
Eqs (42,43) of \cite{laihotwo}.

Note that when $N=3$, the LEC's are those of the full theory, but
it is still necessary to vary the (degenerate) sea and valence
masses independently in order to get all of the needed constants.
Also note that for the $\Delta I=1/2$ amplitude, Eq (2), the
operator subtraction has been performed. For details, see
\cite{laihotwo}.  Finally, for additional redundancy in the
determination of the (8,8), NLO LEC's, one can use nondegenerate
valence quark masses in the $K\to\pi$ amplitudes.

\section{PQS vs PQN}

According to \cite{goltthree}, there are at least two ways of
embedding the left-right QCD penguins into the partially quenched
theory.  We have dubbed these the PQS and PQN methods, based on
the notation in \cite{goltthree}.  For the PQS method one assumes
the right handed part of the penguin operator transforms as a
singlet under the extended symmetry group of the penguin operator;
hence the name partially quenched singlet (PQS) method.  In the
PQN method, the right handed part of the QCD penguin contains only
valence quarks, so that the PQChPT at leading order becomes the
sum of two operators, one of which is just the singlet operator of
the PQS method, while the other does not transform as a singlet
under the extended symmetry group; hence the name partially
quenched non-singlet (PQN) method.

For the left-right QCD penguins we have considered only the PQS
method to NLO because the PQN method requires a two-loop
calculation (for insertions of the non-singlet operator) to the
same order. Only in the full theory do the two methods coincide,
and the contributions of the non-singlet operator vanish.  Even if
one believes that the PQN method is closer to the full theory
because it keeps ``eye diagram" contractions that would be
discarded in the PQS method (see the discussion in
\cite{laihotwo}), it would still be useful to implement the PQS
method, since this method can be carried out to \emph{NLO}.  Of
course, an $N=3$ calculation would resolve the ambiguity.

\section{(8,1), $K\to\pi\pi$ amplitudes}

The ingredients necessary to construct (8,1), $K\to\pi\pi$
amplitudes to NLO are $K\to0$, $K\to\pi$ and $K\to\pi\pi$ at UKX,
all of which require nondegenerate quark masses but no 3-momentum
insertion.  For a detailed description of how to obtain all of the
necessary constants from the various amplitudes, see Section 8 of
\cite{laihotwo}.  Here, a brief description of the operator
subtraction is given.  The power divergence in the $\Delta I=1/2$
matrix elements reduces to an effective quark bilinear times a
momentum independent coefficient \cite{blum},

\begin{equation}
\Theta^{(3,\overline{3})}\equiv \overline{s}(1-\gamma_5)d=
\alpha^{(3,\overline{3})}\textrm{Tr}(\lambda_6\Sigma),
\end{equation}

\noindent to lowest order in ChPT.  The subtraction makes use of
the ratio of $K\to0$ amplitudes, which to NLO is

\begin{eqnarray}
\frac{\langle 0|{\cal O}^{(8,1)}|K^{0}\rangle}{\langle
0|\Theta^{(3,\overline{3})}|K^{0}\rangle} &=&
2\frac{\alpha_2}{\alpha^{(3,\overline{3})}}(m^2_K-m^2_\pi)
\nonumber \\ && +2\frac{\alpha_1}{\alpha^{(3,\overline{3})}}(logs)
+\frac{4}{\alpha^{(3,\overline{3})}}(m^2_K \nonumber \\
&& -m^2_\pi) \left[2\left(e^r_{1,rot}-e^r_{5,rot}\right)m^2_K
 \right.\nonumber \\ &&  \left.
+\left(e^r_{2,rot}\right)Nm^2_{SS}\right].\nonumber
\\ &&
\end{eqnarray}

The subscript, ``\emph{rot}," accounts for the effect of the
subtraction on the (8,1) LEC's to NLO.  They are transformed to
new linear combinations involving the Gasser-Leutwyler
coefficients \cite{kambor}, with the power divergences removed.

Using the ratio determined from the $K\to0$ amplitudes, the power
divergences can be subtracted from $K\to\pi\pi$ amplitudes using
the following formula,

\bea \langle \pi^+\pi^-|{\cal O}^{(8,1)}_{sub}|K^{0}\rangle
&\equiv & \langle \pi^+\pi^-|{\cal O}^{(8,1)}|K^{0}\rangle
\nonumber
\\ && - 2\frac{\alpha_2
}{\alpha^{(3,\overline{3})}}(m^2_K-m^2_\pi) \nonumber \\ && \times
\langle \pi^+\pi^-|\Theta^{(3,\overline{3})}|K^{0}\rangle. \eea

\noindent From the leading order expression for the
$\Theta^{(3,\overline{3})}$ operator,

\bea \langle \pi^+\pi^-|\Theta^{(3,\overline{3})}|K^{0}\rangle=
\frac{i}{f^3}\alpha^{(3,\overline{3})}\frac{2m_\pi-m_K}{m_K-m_\pi},
\eea

\noindent one can see there is a problem when $m_K=m_\pi$ (UK1)
where the amplitudes blows up, even though the final subtracted
amplitude, Eq (5), is finite at UK1.

\section{Enhanced Finite Volume Corrections}

Lin, \etal, \cite{linthree} have done a calculation of the
relevant finite volume Euclidean correlation functions for
$K\to\pi\pi$. They have discovered that the case of the $\Delta
I=1/2$, $K\to\pi\pi$ amplitudes at degenerate quark masses (UK1)
has difficulties in the full theory, and is intractable in the
partially quenched theory because of enhanced finite volume
effects, even at the special kinematics, $m_{sea}=m_{u,d}$.
According to the results of \cite{linthree}, these difficulties do
not afflict the more general kinematics of UKX for $m_K$ strictly
greater than $m_\pi$ in the two flavor dynamical case (at the
special value of $m_{sea}=m_{u,d}$).  In our revised paper we show
that one could use this set of kinematics points instead of UK1 to
obtain all of the NLO LEC's in the two flavor theory.  It was also
pointed out by \cite{linthree} that for $\Delta I=1/2$,
$K\to\pi\pi$ amplitudes in the three flavor dynamical case, in
order to avoid enhanced finite volume effects, one must work in
the full theory.

\bigskip

This research is supported by US DOE Contract No.\
DE-AC02-98CH10886.

\end{document}